\documentclass[print,trackchanges]{ati}	%%two-column, with linenumbers and trackchanges -- 推荐投稿采用以方便排版 
\usepackage{graphicx,times}
\usepackage{lmodern}
\usepackage[sort&compress, numbers,super]{natbib} %%comment this line if not using superscript number citation as [1,6]
\bibpunct[,]{[}{]}{,}{n}{}{,}			%%comment this line for non-numbering citations 'author-year'
\usepackage{url}
\usepackage[colorlinks=true,breaklinks=true, linkcolor=red,urlcolor=magenta,citecolor=blue,anchorcolor=green]{hyperref}
\usepackage{caption}
\usepackage{rotating}
\usepackage{CJKutf8}	%%for inclusion of Chinese characters
\usepackage{fancyhdr}
\usepackage{float}  
\usepackage{placeins}
\usepackage{hyperref}
\usepackage{xurl}
\usepackage{xcolor}
\usepackage{booktabs} % penghao
\usepackage{siunitx} % penghao
\usepackage{multirow} % penghao
\renewcommand{\citet}[1]{\citeauthor{#1}(\citeyear{#1})\cite{#1}}	%%revised \citet{key} command
\begin{document}

   \title{
   {A Python-Based Peeling Framework for Radio Interferometry: Application to uGMRT 650\,MHz Imaging}
}

%   \subtitle{Improving uGMRT 650\,MHz Imaging via Python-Based Peeling}

%%
%% -------------------- Authors and institutes go below -----------------
%% Here is an example of four authors come from three different institutes.
%% For single author or all the authors from one institute, use "\inst{}" only, and comment "\and" lines
   \author{Hao Peng
      \inst{1,2}
      \ORCID{0009-0001-5497-7787}		%ORCID is supported: add \ORCID{} right below \inst{}
   \and Fangxia An\correspondingAuthor{}
     \inst{3}
     \ORCID{0000-0001-7943-0166}
   \and Yuheng Zhang
      \inst{4,5}
      \ORCID{0000-0001-5757-5719}
   \and Srikrishna Sekhar  	%%this shows how to print Chinese
      \inst{6,7}
      \ORCID{0000-0002-8418-9001}
   \and Russ Taylor  	%%this shows how to print Chinese
      \inst{7,8}
      \ORCID{0000-0001-9885-0676}
   \and Xianzhong Zheng  	%%this shows how to print Chinese
      \inst{9}
      \ORCID{0000-0003-3728-9912}
   \and Yongming Liang
      \inst{10,11}
      \ORCID{0000-0002-2725-302X}
   }
%%corresponding author: the person who is responsible for communicating with the journal on behalf of all of the authors of a research paper. This includes tasks such as responding to editorial queries, submitting revisions, and tracking the publication process. 
%% Please give the correspondent's full name (the same as that given in the authors list) and Email address below:
\correspondent{Fangxia An}	%% = corresponding Author
\correspondentEmail{anfangxia@ynao.ac.cn}

\institute{Purple Mountain Observatory, Chinese Academy of Sciences, Nanjing 210046, P. R. China;
          \and % 2
              School of Astronomy and Space Sciences, University of Science and Technology of China, Anhui 230026, P. R. China;
          \and % 3
              Yunnan Observatories, Chinese Academy of Sciences, Yunnan 650217, P. R. China;
          \and % 4
              School of Astronomy and Space Science, Nanjing University, Nanjing, Jiangsu 210093, China
          \and % 5
              Key Laboratory of Modern Astronomy and Astrophysics (Nanjing University), Ministry of Education, Nanjing 210093, China
          \and % 6
              National Radio Astronomy Observatory, NRAO, 1003 Lopezville Road, Socorro, NM 87801, USA
          \and % 7
              Inter-University Institute for Data Intensive Astronomy, Department of Astronomy, University of Cape Town, Private Bag X3, Rondebosch, 7701, Cape Town, South Africa
          \and % 8
              Inter-University Institute for Data Intensive Astronomy, Department of Physics and Astronomy, University of the Western Cape, Robert Sobukwe Road, Bellville, 7535, Cape Town, South Africa
          \and % 9
              Tsung-Dao Lee Institute and State Key Laboratory of Dark Matter Physics, Shanghai Jiao Tong University, Shanghai, 201210, China
          \and % 10
              Institute for Cosmic Ray Research, The University of Tokyo, 5-1-5 Kashiwanoha, Kashiwa, Chiba 277-8582, Japan
          \and % 11
              National Astronomical Observatory of Japan, 2-21-1 Osawa, Mitaka, Tokyo 181-8588, Japan
   }
   \date{Received:~February 04, 2026;   Accepted:April 22, 2026;  Published Online:~XX, 2026; 
   \DOI{ati2026009} }			%% preserved for editors: DOI tail part for each paper
   \citeinfo {Peng, H. et al.}\volume{X}\issue{X} \pages{xx--xx}	%% preserved for editors  
   \StartPage{1} 			%% preserved for editors: starting page number of this paper
   \MonthIssue{April}		%% preserved for editors
   \copyrights {2026}     		%% preserved for editors: =year
%   \manuscriptno{A106}			%editor assigned reference number
%%
   \abstract{
Modern radio interferometric arrays offer high sensitivity, wide fields of view, and broad frequency coverage, but also pose significant data calibration challenges. Standard direction-independent calibration is insufficient to correct direction-dependent effects, such as ionospheric phase distortions and primary beam variations, which produce strong artifacts around bright sources and limit achievable image dynamic range. Built on standard CASA (Common Astronomy Software Applications package) tasks, we present a Python-based direction-dependent calibration and \texttt{peeling} framework, demonstrated using radio continuum imaging data from the upgraded Giant Metrewave Radio Telescope (uGMRT). The framework efficiently subtracts bright-source models and suppresses their associated direction-dependent artifacts, producing significantly flatten backgrounds and improving image fidelity and faint-source detectability. We further introduce an optimized ``model-restoration'' strategy that mitigates direction-dependent artifacts while preserving the overall flux densities of bright sources that are themselves of scientific interest. For fields containing multiple bright sources, sequential application of the framework systematically reduces background noise, thereby increasing sensitivity and faint-source detectability. The framework is Python-based, CASA-compatible, and readily applicable to other mid- and low-frequency interferometric arrays, with the code publicly released alongside this paper. 
  }
%% Provide at most 8 Keywords within 'abstract{}' environment. ATI uses Unified Astronomy Thesaurus concepts as AAS journals. 
%% Please find your keywords by visiting https://astrothesaurus.org/concept-select/   or select from the journal's web http://www.ati.ac.cn

   \authorrunning{ASTRONOMICAL TECHNIQUES \& INSTRUMENTS }   %% author_head on even pages 
   \titlerunning{Peng Hao. et al.: ~Prepare a LaTeX Manuscript for ATI }  %% short authors_title head on odd pages
   \maketitle
   \setcounter{page}{\Page}	%% preserved for editors: starting page 
%
%____________________________________________________ sections below
%
\section{Introduction}
\label{sect:intro}

Radio interferometry significantly enhances the spatial resolution of radio observations through the technique of aperture synthesis\textsuperscript{\citep{Thompson2017}}, enabling resolution comparable to that achieved at optical wavelengths. Modern radio interferometric arrays, such as the LOw Frequency ARray (LOFAR\textsuperscript{\citep{van_Haarlem2013}}) and the Karl G. Jansky Very Large Array (VLA), routinely achieve sub-arcsecond angular resolution\textsuperscript{\citep{Dzib2025}}, which is crucial for resolving the morphology and internal structure of astrophysical sources. In parallel, significant advances in receiver technology, correlator performance, and survey strategies have led to dramatic improvements in the sensitivity and survey efficiency of modern radio interferometric arrays\textsuperscript{\citep{Best2023}}. Together, these developments have established radio continuum observations as a powerful tool for a broad range of astrophysical studies. 

Despite these advances, interferometric imaging is fundamentally limited by incomplete sampling of the spatial-frequency (\emph{uv}) plane, as determined by the number of antennas, the array configuration, and Earth-rotation synthesis. Consequently, owing to the lack of short and zero spacings, interferometers resolve out large-scale extended structures that are readily accessible to single-dish telescopes, and the resulting images are intrinsically limited in dynamic range by residual sidelobes and calibration errors. In addition, the raw visibilities measured by an interferometer encode both the sky signal and systematic instrumental and atmospheric responses, and therefore cannot be directly interpreted as the intrinsic sky brightness distribution.
% The Fourier transform of the measured visibilities yields a \emph{dirty image} with pronounced sidelobe structures, requiring careful calibration and imaging procedures to correct instrumental, atmospheric, and observational effects and to recover a physically meaningful sky brightness distribution.

To correct these systematic effects, standard radio interferometric data reduction employs direction-independent calibration, which aims to correct instrumental and atmospheric response errors that are assumed to be uniform across the field of view. These include absolute flux-density scaling, instrumental delay calibration, bandpass calibration, complex gain calibration, and self-calibration. Together, these steps account for time- and frequency-dependent instrumental responses, as well as first-order atmospheric effects\textsuperscript{\citep{Thompson2017}}, significantly reducing residual phase and amplitude errors—such as those induced by large-scale atmospheric fluctuations\textsuperscript{\citep{Perley1999}}, and thereby improving image fidelity and dynamic range.

However, direction-independent calibration is intrinsically limited when gain 
errors vary across the field of view. With the increasing sensitivity, wide field of view, and broad frequency coverage of modern radio interference arrays, direction-dependent effects have become increasingly significant. In particular, the gain variations introduced by the antenna primary beam are direction-dependent across the field of view\textsuperscript{\citep{Smirnov2011}}, whereas propagation effects, such as ionospheric electron density fluctuations at low frequencies and tropospheric water vapor at high frequencies, introduce excess path-length variations along different lines of sight, resulting in line-of-sight–dependent phase distortions\textsuperscript{\citep{Thompson2017}}. If left uncorrected, these effects give rise to stripe-like or radial artifacts around bright sources, significantly limiting image dynamic range and degrading the accuracy of source detection and photometry. 

To mitigate these limitations, direction-dependent calibration techniques, most notably ``peeling", have been developed and applied for over two decades\textsuperscript{\citep{Noordam2004,Intema2009,Retana2025}}. \texttt{Peeling} isolates the signal of a bright source and solves independently for its direction-dependent complex gains, thereby correcting phase and amplitude errors along the corresponding line of sight and substantially reducing the prominent artifacts surrounding the target bright source\textsuperscript{\citep{Intema2009}}. Both sequential \texttt{peeling} of individual sources and simultaneous calibration of multiple sources have been shown to be effective in mitigating direction-dependent gain errors\textsuperscript{\citep{Cotton2021}}.

Numerous \texttt{peeling}-based approaches have been proposed, differing in their implementation and scope. Some methods improve image fidelity by subtracting problematic bright sources from the visibilities or from the final images\textsuperscript{\citep{Williams2019}}. Other techniques, such as the SPAM (Source Peeling and Atmospheric Modeling) framework\textsuperscript{\citep{Intema2009}}, were developed primarily for low-frequency radio interferometric observations (e.g., VLA at 74\,MHz and GMRT 610\,MHz) and mitigate direction-dependent errors by modeling ionospheric phase distortions using multiple peeled sources. However, such strategies are unsuitable when bright sources are of scientific interest, or may not fully account for other direction-dependent effects, such as antenna primary beam variations.

In this work, we develop a Python-based, CASA-compatible \texttt{peeling} strategy capable of suppressing direction-dependent artifacts while preserving bright sources as scientifically valuable targets. Its performance is demonstrated using radio continuum imaging data from the uGMRT, whose observation often exhibit pronounced imaging artifacts, arising from the relatively sparse sampling of long baselines and the enhanced impact of ionospheric phase distortions at low radio frequencies\textsuperscript{\citep{Intema2009}}.

The structure of this paper is as follows. $\S$~\ref{sect:data} describes the uGMRT observations and the standard direction-independent calibration procedures. The \texttt{peeling} methodology and the optimized model-restoration strategy are introduced in $\S$~\ref{sect:peeling}. We then discuss and summarize the main results and implications in $\S$~\ref{sect:discussion} and $\S$~\ref{sect:conclusions}.

\section{Observations and standard Data reduction}
\label{sect:data}

\subsection{Observations}
The radio continuum data used in this work were obtained with the uGMRT (proposal code: 41\_103, PI: Fangxia An) in Band 4, covering the frequency range 550–850\,MHz. The target field, BOSS J0210+0052 (hereafter J0210; RA = 02:10:00.00, Dec = +00:52:00.00), lies within the SDSS Stripe 82 region. This field was selected because it hosts both a prominent galaxy overdensity and a quasar overdensity at redshift $z\sim2.2$, identified by the MApping the Most Massive Overdensity Through Hydrogen (MAMMOTH) project\textsuperscript{\cite{Cai2016,Cai2017,Zheng2021,Zhang2022,Shi2021}} 
together with its narrow-band follow-up observations using the Subaru Hyper Suprime-Cam (HSC), conducted as part of the MAMMOTH–Subaru project\textsuperscript{\citep{Liang2021,Liang2025}}.

The primary scientific goal of our uGMRT observations is to identify dust-unbiased members of this high-redshift galaxy and quasar overdensity. This population is largely inaccessible to the MAMMOTH–Subaru survey, which primarily targets Ly$\alpha$ emitters (LAEs) and is therefore strongly biased against dusty systems due to the sensitivity of Ly$\alpha$ emission to dust attenuation.

To fully cover the LAE-traced overdense region, we designed three overlapping uGMRT Band\,4 pointings: J0210-1 (RA = 02:11:18.00, DEC = + 00:48:00.00), J0210-2 (RA = 02:09:48.00, DEC = + 00:54:00.00), and J0210-3 (RA = 02:08:06.00, DEC = +01:00:00.00). Each pointing was observed over two long tracks of 6 and 7 hours, respectively, resulting in a total integration time of 13 hours per pointing and 39\,hours for the entire J0210 field.

During each observation track, standard flux density and bandpass calibrators 3C\,147 or 3C\,48 were observed for $\sim$10\,minutes at the start and the end of each track, while the phase calibrator 0204+152 was observed for 5\,minutes every $\sim$30\,minutes to track time-dependent complex gain variations. At the effective observing frequency of $\sim$650\,MHz, the uGMRT Band\,4 primary beam has a full width at half maximum of $\sim$38\,arcmin.

\subsection{Standard direction-independent calibration}
\label{sect:DIC}

We performed the initial calibration of the raw uGMRT visibilities, which were converted into CASA-compatible measurement sets for data reduction, for each individual pointing using the Common Astronomy Software Applications package (CASA, version 6.6.4.34;\textsuperscript{\citep{CASATeam2022}}), following the standard direction-independent calibration procedures described in the GMRT tutorials\footnote{\url{https://gmrt-tutorials.readthedocs.io/en/latest/index.html}}.

\begin{figure}%[ht!]
    \begin{minipage}[t]{0.99\linewidth}  %% use \textwidth for full-page width crossing two columns
    \centering
    \includegraphics[width=0.85\textwidth,angle=0,scale=1]{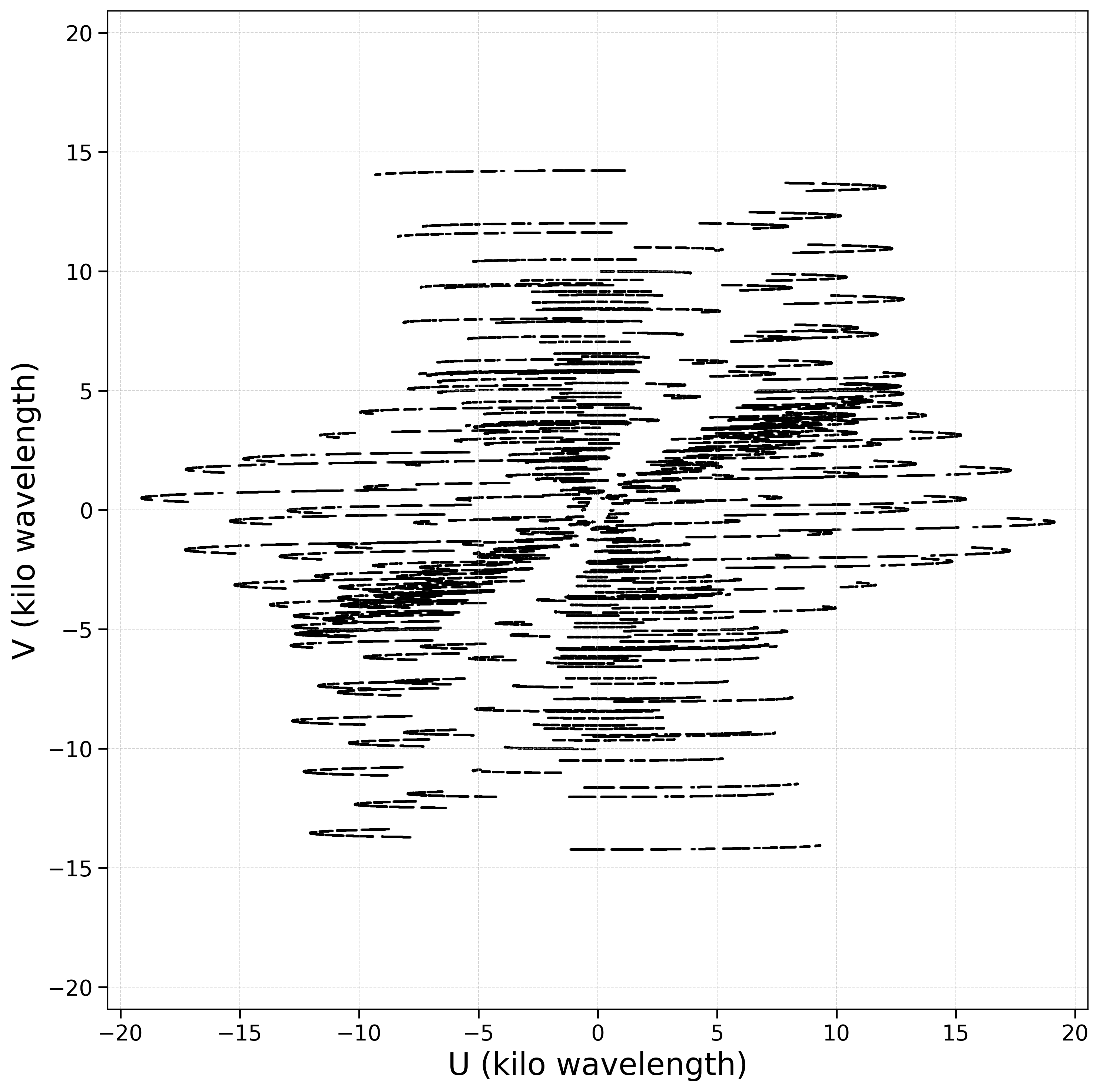}
    \caption{The uv-coverage of the J0210-3 subfield. Each point represents one baseline visibility in the (u, v) plane.}
    \label{fig:3+4_uv}
    \end{minipage}
\end{figure}

Briefly, the standard direction-independent calibration and imaging workflow consists of the following steps: 
\begin{enumerate}
    \item {\bf Initial flagging and data preparation:} Owing to reduced sensitivity, unstable bandpass response, and increased susceptibility to radio-frequency interference (RFI) near the band edges, we first discarded the outermost 3\% of frequency channels at both the low- and high-frequency ends of the raw visibilities using the \texttt{flagdata} task. We then employed the \texttt{plotms} task to inspect amplitude–time and amplitude–frequency distributions for each antenna and scan, allowing us to identify malfunctioning antennas, anomalous time ranges, and other problematic data. These data were then flagged manually using \texttt{flagdata}. In addition, we applied automated RFI excision using the \texttt{tfcrop} mode of \texttt{flagdata} to remove outliers before the initial calibration. 
    \item {\bf Initial calibration (delay, gain, and bandpass):} The primary flux-density and bandpass calibrator (3C\,147 or 3C\,48) was used to establish the initial calibration. Delay, complex gain, and bandpass solutions were derived sequentially using the \texttt{gaincal} and \texttt{bandpass} tasks, producing the corresponding calibration tables. These calibration solutions were then applied to the target-field visibilities using the \texttt{applycal} task. 
    \item {\bf Automated and additional manual flagging:} To further improve data quality, additional automated RFI flagging was performed using the \texttt{rflag} mode of the \texttt{flagdata} task. After all problematic data had been flagged, we used \texttt{clearcal} to re-initialize the calibration for visibilities and repeated the standard calibration procedures described above. This approach ensured that the final calibration tables were derived from a less-biased dataset. The re-calibrated visibilities were subsequently inspected again using \texttt{plotms}, revealing residual contamination near the band edges. Consequently, the first and last 4\% of frequency channels were further excluded. After this additional flagging, the effective frequency range retained for subsequent analysis spans 565--737\,MHz, consistent with the frequency coverage adopted in previous uGMRT Band\,4 continuum studies, particularly those based on deep survey observations\textsuperscript{\citep{Lal2025}}. 
    \item {\bf Self-calibration:} To further mitigate residual antenna-based complex gain errors and enhance image fidelity and dynamic range, we applied self-calibration using the automated pipeline CAsa Pipeline-cum-Toolkit for Upgraded Giant Metrewave Radio Telescope data REduction (CAPTURE\textsuperscript{\citep{Kale2021}}). Adopting the recommended workflow and parameter settings provided in the CAPTURE tutorials, we performed four iterations of phase-only self-calibration, followed by four iterations of combined amplitude-and-phase self-calibration. The solution interval (\texttt{solint}) was progressively shortened from 8.0\,min in the initial cycle to 1.0\,min in the final cycle. The uv-coverage of the J0210-3 subfield is shown in Fig.~\ref{fig:3+4_uv}, illustrating the baseline sampling that supports the imaging and self-calibration. This subfield is used for the global peeling analysis in $\S$~\ref{sect:global_impact}. 
    \item {\bf Imaging:} Following each self-calibration iteration, imaging was performed using the \texttt{tclean} task in CASA. The image obtained after the final self-calibration iteration corresponds to the optimal result achievable with direction-independent calibration and is commonly adopted as the final science image in most uGMRT radio continuum studies\textsuperscript{\citep[e.g.,][]{Dokara2023, Sinha2023, Lal2025}}.
\end{enumerate}

%%%%#### Figure 1 #####

\begin{figure*}%[ht!]
    \begin{minipage}[t]{0.999\linewidth}  %% use \textwidth for full-page width crossing two columns
    \centering
    \includegraphics[width=0.9\textwidth,angle=0,scale=1]{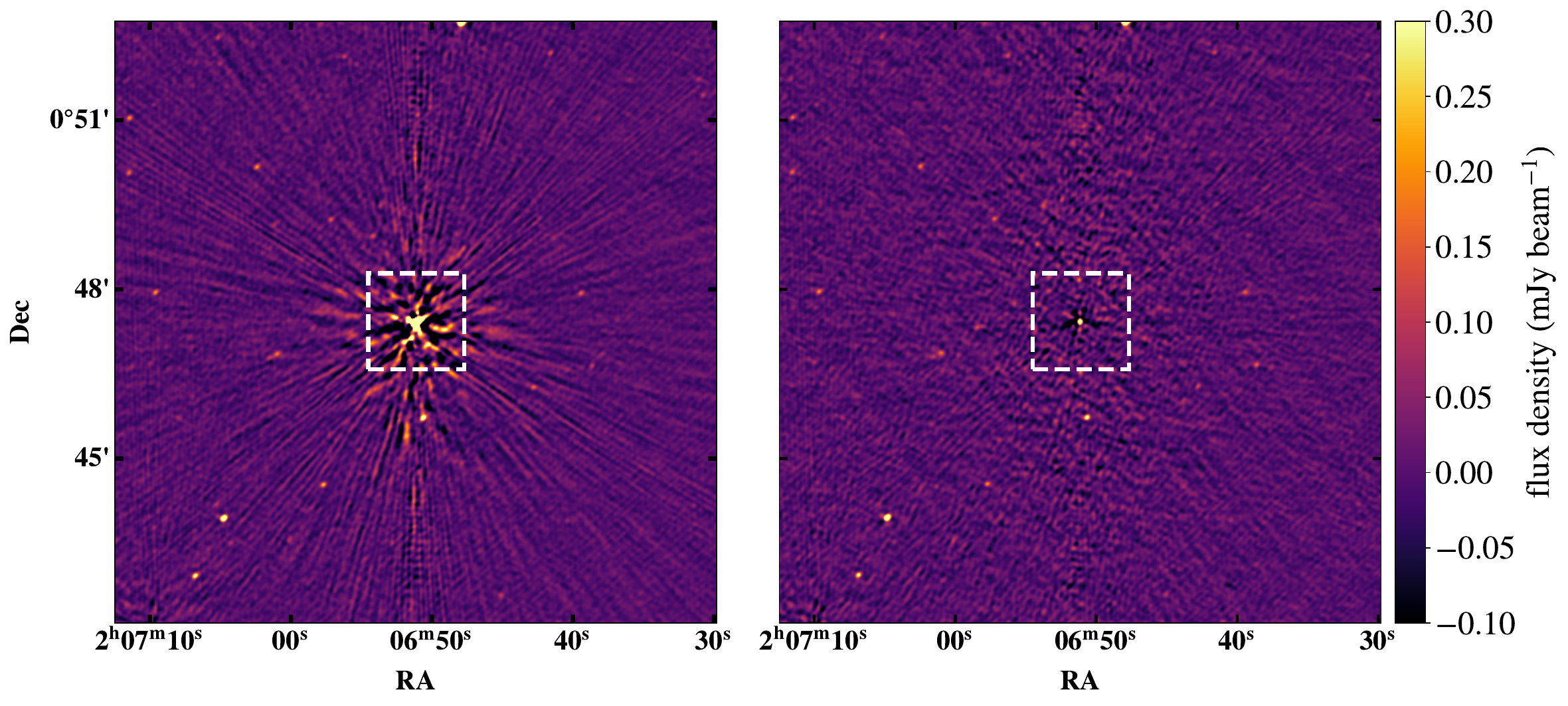}
    \caption{Comparison illustrating the effect of the \texttt{peeling} procedure on the uGMRT 650\,MHz imaging data. The white dashed box outlines the \texttt{peeling} mask region. The left panel shows a sub-region of the image obtained after the eighth (final) round of self-calibration, representing the optimized result of direction-independent calibration. The right panel presents the corresponding image after applying direction-dependent calibration and model subtraction to the target bright source, in which artifacts associated with the bright source are substantially suppressed, resulting in a significantly flatter background and enabling neighboring faint sources to be more clearly discerned.}
    \label{fig:3+4_peeling}
    \end{minipage}
\end{figure*}

However, as illustrated in the left panel of Fig.~\ref{fig:3+4_peeling}, images produced using standard direction-independent calibration still exhibit prominent, radially distributed artifacts around bright sources. These artifacts severely hinder the accurate detection and photometry of both the bright sources themselves and nearby faint sources, and can degrade the overall image dynamic range\textsuperscript{\citep{Williams2019,Salunkhe2025}}. Such artifacts primarily arise from uncorrected direction-dependent gain variations caused by spatially varying antenna primary beam responses and ionospheric phase fluctuations. To address these limitations, we develop a direction-dependent calibration scheme specifically optimized for uGMRT radio continuum data, as described in the following section.

%%%%#### Figure 2 #####

\section{Peeling}
\label{sect:peeling}
\texttt{Peeling} is a widely used direction-dependent calibration technique in radio interferometry, in which calibration solutions are derived for specific directions within the field of view to correct localized gain errors. As described in $\S$~\ref{sect:DIC}, applying \texttt{peeling} is crucial for improving the image fidelity and dynamic range of our uGMRT Band\,4 imaging data by suppressing prominent artifacts around strong radio sources caused by direction-dependent gain effects (the left panel of Fig.~\ref{fig:3+4_peeling}). 

Our \texttt{peeling} procedure is applied to the visibility data after completion of the eighth (and final) round of self-calibration. Given that the problematic bright sources in this field are unlikely to be physically associated with the high-redshift protocluster that is the primary science target of our uGMRT observations, we first develop a standard \texttt{peeling} scheme in which direction-dependent gain solutions are solved toward each bright source and the corresponding source model is subsequently subtracted from the visibilities (see $\S$~\ref{sect:standard_peeling}).

To ensure the broader applicability of our \texttt{peeling} framework, we also consider scenarios in which the bright sources themselves are of scientific interest. In this case, we implement an alternative scheme in which the bright-source model is restored to the visibilities after direction-dependent calibration ($\S$~\ref{sect:sources_retain}).

Finally, for fields containing multiple bright radio sources, we apply the \texttt{peeling} procedure sequentially to individual sources and assess the cumulative impact on image quality and sensitivity, as presented in $\S$~\ref{sect:global_impact}.

\subsection{Sources to remove}
\label{sect:standard_peeling}

The J0210 field contains several bright radio sources whose uncorrected direction-dependent effects dominate the local noise in their vicinity. Although these sources are of intrinsic astrophysical interest, they are not the primary science targets of our uGMRT observation and significantly degrade the fidelity of the surrounding regions through time- and direction-dependent gain variations\textsuperscript{\citep{Bhatnagar2008}}. 

We therefore develop our modified \texttt{peeling} strategy, following the standard \texttt{peeling} framework described by \citeauthor{Williams2019} \citeyear{Williams2019}\textsuperscript{\citep{Williams2019}}, in which direction-dependent gain solutions are derived independently toward each problematic bright source. These source-specific gain solutions are applied to the visibilities, after which the corresponding bright-source model is subtracted from the data. This procedure effectively reduces residual artifacts and improves the detectability and photometric accuracy of nearby faint sources within the affected regions.

\begin{figure*}[ht!]
    \begin{minipage}[t]{0.999\linewidth}  %% use \textwidth for full-page width crossing two columns
    \centering
    \includegraphics[width=0.99\textwidth,angle=0,scale=1]{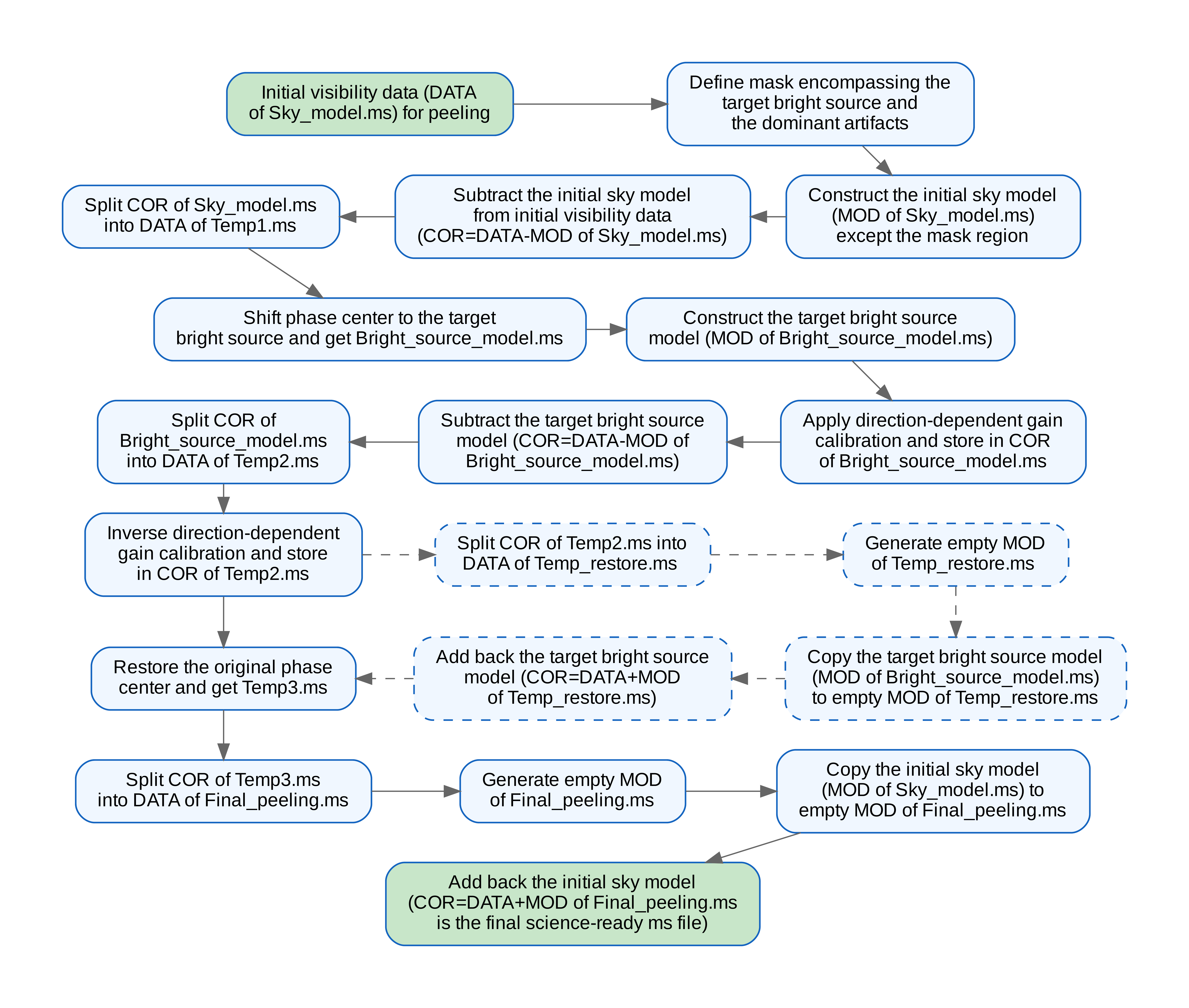} %.png/jpg/pdf/eps extension is not required
    \caption{Schematic overview of the direction-dependent calibration framework adopted in this work. Solid boxes represent the standard \texttt{peeling} procedure, in which the model of a problematic bright source is subtracted from the visibility data after direction-dependent calibration. Dashed boxes highlight the additional steps introduced in our optimized model-restoration strategy, designed to preserve the flux density and morphology of bright sources of scientific interest. All TempX.ms are deleted immediately after use to minimize storage usage. DATA, MOD and COR represent DATA column, MODEL\_DATA column, and  CORRECTED\_DATA column respectively.}
    \label{fig:peeling_flowchart}
    \end{minipage}
\end{figure*}

The overall architecture of our \texttt{peeling} workflow is illustrated in Fig.~\ref{fig:peeling_flowchart}. The step-by-step Python-based workflow implemented within the CASA environment to perform this direction-dependent calibration and source subtraction as follows. We note that the phase-center shifting step (Step~\ref{item:step3}) is optional. Nevertheless, centering the bright source typically improves the stability of the calibration, as the visibility phases can be more naturally assumed to be near zero. 

%%%%#### Figure 3 #####

\begin{enumerate} %%advanced function please use enumitem or enumerate packages
  % 1
  \item \label{item:step1} \textbf{Preparation of visibilities and definition of the \texttt{peeling} mask.} 
  The measurement set file obtained after the eighth (final) round of self-calibration was designated as \textbf{Sky\_model.ms} and adopted as the input dataset for the \texttt{peeling} procedure, with the \texttt{DATA} column providing the initial visibilities.
  
  To isolate each problematic bright source for direction-dependent calibration, a spatial mask centered on the source was defined using the CASA Region Text Format (CRTF). The mask size was chosen to be slightly larger than the intrinsic extent of the bright source in order to encompass both the source emission and the dominant surrounding artifacts. In this work, a uniform square mask of $128 \times 128$ pixels ($102.4'' \times 102.4''$) was adopted for all peeled sources. In practice, the Python implementation allows the mask shape and size to be flexibly adjusted based on the morphology of individual sources. 
  % 2
  \item \label{item:step2} \textbf{Construction and subtraction of the initial sky model.} An initial sky model was constructed using the \texttt{tclean} task applied to \textbf{Sky\_model.ms}, with the mask defined in Step~\ref{item:step1} used to exclude the region containing the target bright source during deconvolution. As a result, the derived model represents all significant emission outside the masked region and was written to the \texttt{MODEL\_DATA} column of \textbf{Sky\_model.ms}.

  This sky model was then subtracted from the original visibilities using the \texttt{uvsub} task, producing residual visibilities stored in the \texttt{CORRECTED\_DATA} column of \textbf{Sky\_model.ms}. These residuals were then written to a new measurement set file, \textbf{Temp1.ms}, by copying the \texttt{CORRECTED\_DATA} column of \textbf{Sky\_model.ms} into the \texttt{DATA} column via the \texttt{split} task.
  
  Provided that the initial sky model is sufficiently accurate, i.e., that emission from all sources outside the mask region has been effectively subtracted, the visibilities in the \texttt{DATA} column of \textbf{Temp1.ms} predominantly contain the signal of the target bright source within the mask. 
  % 3
  \item \label{item:step3} \textbf{Phase-center shifting and modeling of the target bright source.} To precisely model the target bright source, we shifted the phase center of the visibilities in the \texttt{DATA} column of \textbf{Temp1.ms} to the source coordinates using the \texttt{phaseshift} task, writing the output to the \texttt{DATA} column of a new measurement set file, \textbf{Bright\_source\_model.ms}. 
  
  We then performed detailed modeling of the target bright source using the \texttt{tclean} task. The deconvolution threshold (key parameters: threshold=``0.05mJy'', niter=20.) was iteratively adjusted while monitoring the residual images to ensure that the model recovered the majority of the source flux. The final model of the target bright source was stored in the \texttt{MODEL\_DATA} column of \textbf{Bright\_source\_model.ms}.
  
  At this stage, the visibilities in the \texttt{DATA} column of \textbf{Bright\_source\_model.ms} are phase-centered on the target bright source, and the emission within the mask region is dominated by that source.
  
  % 4
  \item \label{item:step4} \textbf{Direction-dependent gain calibration and subtraction of the target bright source.} Using the visibility data and source model contained in \textbf{Bright\_source\_model.ms}, we derived direction-dependent complex gain solutions for the target bright source with the \texttt{gaincal} task (key parameters: solint=``inf'', calmode=``ap'', uvrange=``''(all). Aiming to obtain a high signal-to-noise ratio gain solution and achieve both amplitude and phase calibration for the bright source). Owing to the phase centering and the dominance of the target source within the masked region, the resulting gain solutions are driven primarily by the target bright source and are therefore representative of the instrumental and propagation effects in its direction.

  The derived gain calibration table was then applied to the \texttt{DATA} column of \textbf{Bright\_source\_model.ms} using the \texttt{applycal} task, and the calibrated visibilities were written to the \texttt{CORRECTED\_DATA} column. We then subtracted the source model from the calibrated visibilities using \texttt{uvsub}. The resulting residual visibilities were copied into the \texttt{DATA} column of a new measurement set \textbf{Temp2.ms} with the \texttt{split} task.

At this stage, the \texttt{DATA} column of \textbf{Temp2.ms} comprises the residual visibilities obtained after subtracting both the target bright source within the \texttt{peeling} mask and the initial sky model outside the \texttt{peeling} mask, along with residual background noise. 
  % 5
  \item \label{item:step5} \textbf{Reversing the direction-dependent calibration and restoring the original phase center.} To prevent the direction-dependent gain solutions derived toward the target bright source from affecting signals in other sources, we reversed the previously applied calibration by constructing and applying an ``inverse table". For each antenna-based complex gain solution $ g_i = A_i e^{j\phi_i} $, the inverse gain was computed as $ g_i^{-1} = A_i^{-1} e^{-j\phi_i} $, yielding an inverse gain table with the same structure as the original direction-dependent gain table. 

  This inverse gain table was applied to the \texttt{DATA} column of \textbf{Temp2.ms} using the \texttt{applycal} task, and the resulting visibilities were written to the \texttt{CORRECTED\_DATA} column.

  Subsequently, we restored the original phase center of the observation by applying the \texttt{phaseshift} task to all data columns of \textbf{Temp2.ms}, producing a new measurement set, \textbf{Temp3.ms}. Finally, the corrected visibilities were extracted into \textbf{Final\_peeling.ms} by copying the \texttt{CORRECTED\_DATA} column of \textbf{Temp3.ms} into the \texttt{DATA} column via the \texttt{split} task.
   
  At this stage, the \texttt{DATA} column of \textbf{Final\_peeling.ms} is equivalent to that of \textbf{Temp1.ms} in terms of phase center, background noise, and signals from other sources, with the only difference being that the target bright source has been effectively peeled from the data. % Thereby preparing for the subsequent addition of the initial sky model.
  % 6
  \item \label{item:step6} \textbf{Restoring the initial sky model.} To reconstruct the final dataset containing both the peeled field and the emission from all other sources, we added back the initial sky model to \textbf{Final\_peeling.ms} through the following procedure. First, we invoked the \texttt{tclean} task on \textbf{Final\_peeling.ms} with the parameter \texttt{niter=0}, thereby creating a structurally valid but empty \texttt{MODEL\_DATA} column without performing any deconvolution. 
  
  We then copied the \texttt{MODEL\_DATA} column from \textbf{Sky\_model.ms}, which contains the initial sky model of all sources outside the \texttt{peeling} mask, into the corresponding empty \texttt{MODEL\_DATA} column of \textbf{Final\_peeling.ms}. Finally, we used the \texttt{uvsub} task with the \texttt{reverse=True} option to add this sky model back to the visibilities, effectively performing a ``data plus model" operation and writing the result to the \texttt{CORRECTED\_DATA} column of \textbf{Final\_peeling.ms}.
  
At this stage, the \texttt{CORRECTED\_DATA} column of \textbf{Final\_peeling.ms} contains the restored emission from all sources outside the \texttt{peeling} mask, while the contribution of the target bright source has been removed. The resulting dataset preserves the original phase center and calibration state and constitutes the final science-ready measurement set used in our science analyses.
\end{enumerate}

Fig.~\ref{fig:3+4_peeling} demonstrates the effectiveness of our \texttt{peeling} procedure. Although small residual emission from the central bright source remains in the peeled image, as discussed in $\S$~\ref{sect:4.1}, this reflects the necessary balance between avoiding over-cleaning of neighboring faint sources and achieving sufficient modeling of the target bright source. However, as shown in Fig.~\ref{fig:3+4_peeling}, in the peeled image, both the target bright source and the associated direction-dependent artifacts are effectively suppressed, resulting in a significantly flatter background over an extended region surrounding the target bright source, well beyond the \texttt{peeling} mask. As a consequence, neighboring faint sources that were previously obscured by these artifacts become clearly detectable, with improved image fidelity. 

\begin{figure*}%[ht!]
    \begin{minipage}[t]{0.999\linewidth}  %% use \textwidth for full-page width crossing two columns
    \centering
    \includegraphics[width=0.95\textwidth,angle=0,scale=1]{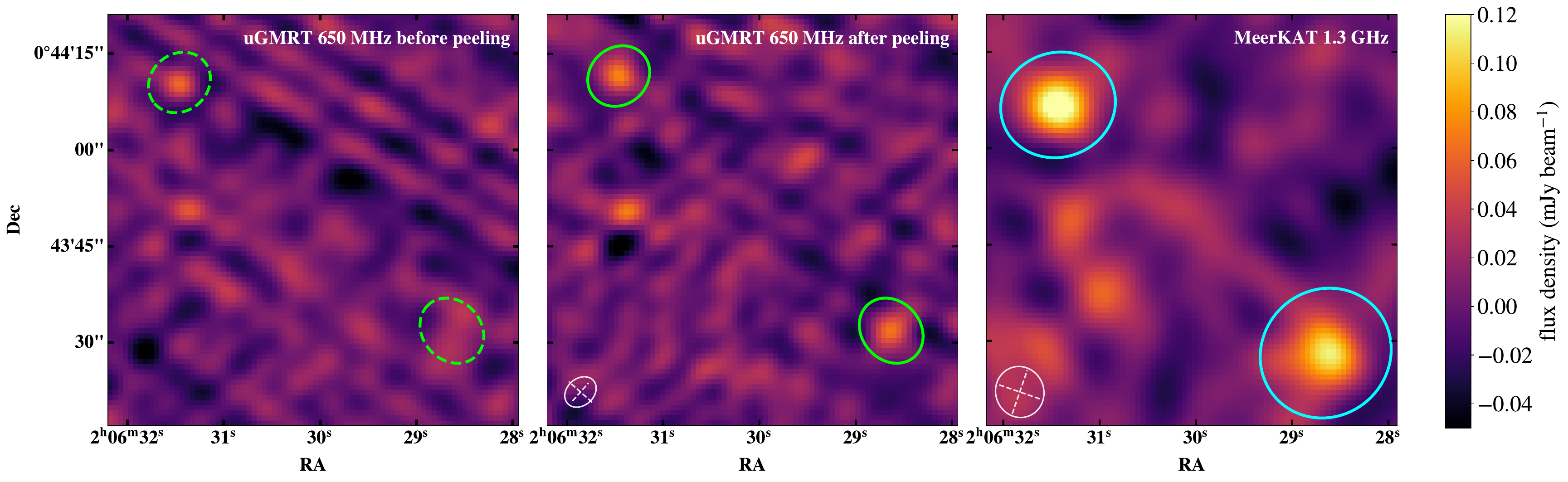} %.png/jpg/pdf/eps extension is not required
    \caption{Comparison of radio source detections in the vicinity of the target bright source before (left panel) and after (middle panel) the \texttt{peeling} procedure, with independent validation from MeerKAT 1.3\,GHz data (right panel). In the post-peeling image (middle panel), two faint radio sources (solid green ellipses) are detected at significance levels exceeding $3\,\sigma$. These two sources are located at angular separations of approximately 350$''$ and 410$''$ from the target bright source.
    These sources are severely obscured by strong direction-dependent artifacts in the direction-independent calibrated image (left panel) and are therefore missed by the source detection algorithm prior to \texttt{peeling}. Their positions, as detected after \texttt{peeling}, are indicated by dashed green ellipses in the left panel. The right panel shows the corresponding MeerKAT 1.3\,GHz reference image, in which the same sources (solid cyan ellipses) are detected at $>5\,\sigma$, thereby independently confirming the reality of these detections. The synthesized beam (indicated by the white line in the bottom-left corner) is $5.4'' \times 4.2''$ for the uGMRT 650\,MHz image and $8.0'' \times 7.5''$ for the MeerKAT 1.3\,GHz image.}
    \label{fig:3+4_comparing_gwb_peeling_MeerKAT}
    \end{minipage}
\end{figure*}

To quantitatively assess the detectability of these faint sources, we performed source extraction using the Python Blob Detection and Source Finder (\texttt{PyBDSF})\textsuperscript{\cite{Mohan2015}}. The same source-finding and RMS (Root Mean Square noise) estimation parameters\footnote {Source-finding thresholds: \texttt {thresh\_isl} = 3.0, \texttt {thresh\_pix} = 3.0; RMS estimation parameters: \texttt {rms\_box} = (40, 13).} were adopted for both the pre- and post-peeling images.
As illustrated in Fig.~\ref{fig:3+4_comparing_gwb_peeling_MeerKAT}, several neighboring faint sources are detected above $3\,\sigma$ only after \texttt{peeling}, primarily due to the suppression of strong direction-dependent artifacts that locally obscure faint emission. Their reality is further corroborated by independent detections in MeerKAT 1.3\,GHz observations. 

%%%%#### Figure 4 #####

\begin{figure}%[ht!]
    \begin{minipage}[t]{0.999\linewidth}  %% use \textwidth for full-page width crossing two columns
    \centering
    \includegraphics[width=0.95\textwidth,angle=0,scale=1]{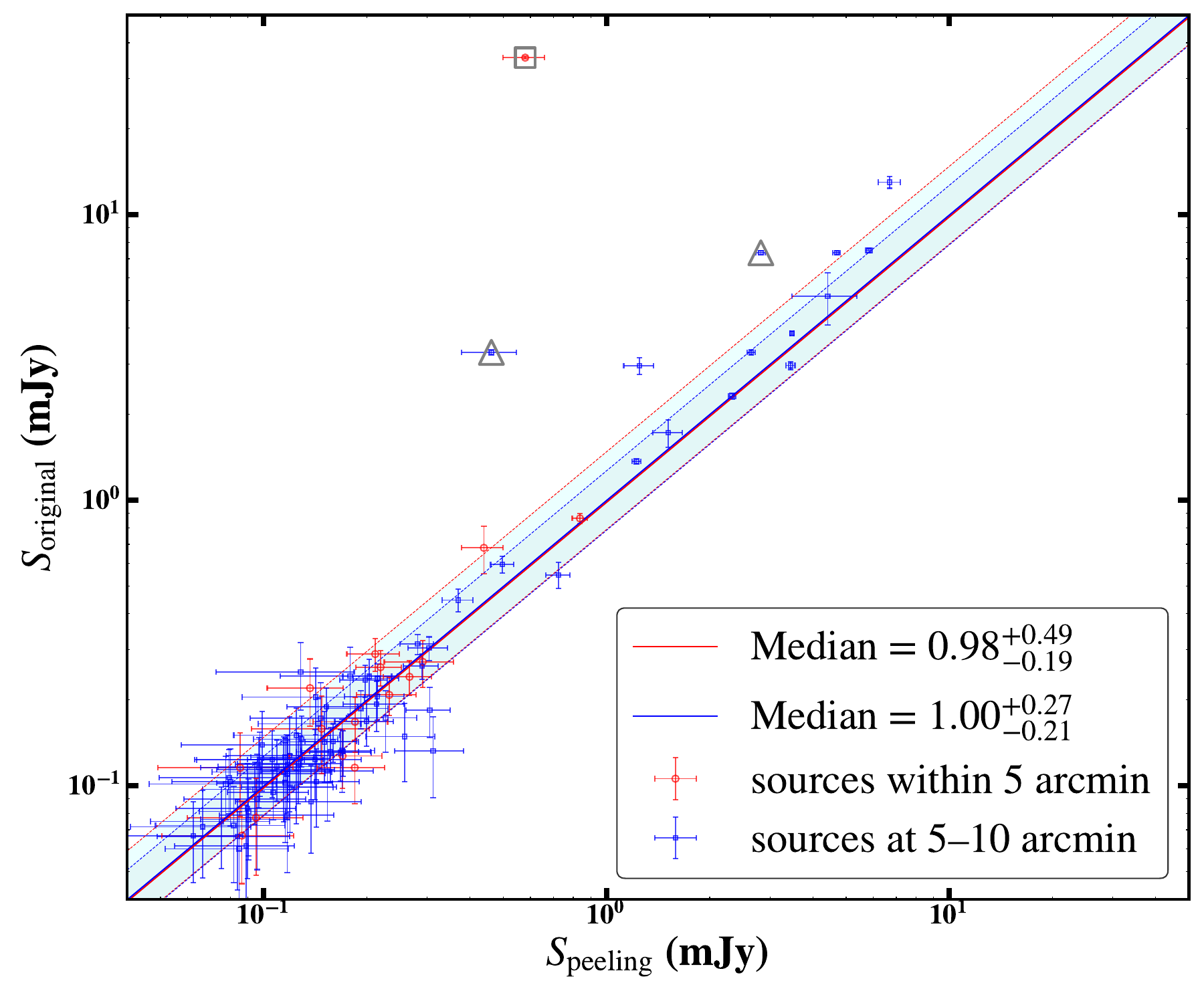}
    \caption{Comparison of source flux densities before and after \texttt{peeling}. Red open circles denote sources located within 5\,arcmin of the target bright source, while  blue open squares indicate sources at radial distance of 5--10\,arcmin. The median ratios of pre-peeling to post-peeling flux densities are \(0.98^{+0.49}_{-0.19}\) and \(1.00^{+0.27}_{-0.21}\) for the inner and outer regions, respectively. The shaded region bounded by dashed lines marks the 16th--84th percentile range (corresponding to the 1$\sigma$ interval).}
    \label{fig:3+4_comparing}
    \end{minipage}
\end{figure}

To evaluate the impact of the \texttt{peeling} procedure on source flux density measurements, we compared the flux densities of identical sources detected before and after \texttt{peeling}. As shown in Fig.~\ref{fig:3+4_comparing}, sources were analyzed in two radial bins relative to the target bright source: within 5~arcmin and  between 5--10~arcmin. The median ratios of pre-peeling to post-peeling flux densities are \(0.98^{+0.49}_{-0.19}\) and \(1.00^{+0.27}_{-0.21}\) for the inner and outer bins, respectively. Both the median values and the overall distribution in Fig.~\ref{fig:3+4_comparing} indicate that, for the majority of faint sources near the target bright source, the measured flux densities remain consistent within uncertainties after \texttt{peeling}. 

However, a small number of outliers contribute to the increased scatter in the flux ratio distribution. One such case, marked by open gray triangles in Fig.~\ref{fig:3+4_comparing}, corresponds to a source that is resolved into two distinct components after \texttt{peeling}, leading to an apparent flux discrepancy. 
Another outlier, indicated by an open black square, corresponds to the target bright source itself. The residual emission arising from imperfect source modeling is detected as a spurious source in the post-peeling image. The origin and implications of this imperfect modeling are discussed in detail in $\S$~\ref{sect:discussion}. 

For the remaining sources, flux density differences that exceed the corresponding measurement uncertainties are primarily attributable to small shifts in the fitted source centroids and the associated extraction regions. 
We compared the source positions in the J0210-3 subfield before and after peeling and found median offsets of 0.01$''$ in RA and 0.00$''$ in Dec, indicating that the \texttt{peeling} process does not introduce any systematic astrometric bias. The remaining small centroid shifts are consistent with random scatter. Our visual inspection further suggests that these shifts are associated with artifacts from the bright source, whose suppression by the \texttt{peeling} process leads to small variations in the fitted source centroids.

%%%%#### Figure 5 #####

\subsection{Sources to retain}
\label{sect:sources_retain}

As outlined in $\S$~\ref{sect:standard_peeling}, the standard \texttt{peeling} procedure effectively removes artifacts associated with the strong bright source and mitigates its influence on surrounding faint sources. 
However, a notable limitation arises when the peeled bright source itself is a scientific target: permanently subtracting its model restricts further analysis of its  intrinsic structure and flux. 

To address this limitation, we optimized the \texttt{peeling} procedure described in $\S$~\ref{sect:standard_peeling} by adding back the target bright source model after the direction-dependent calibration has been applied. This approach preserves the benefits of artifact suppression while retaining the scientifically relevant signal of the target source.

Building upon the standard \texttt{peeling} workflow, our optimized processing pipeline requires only a three-step extension starting from Step~\ref{item:step5} of the procedure described in $\S$~\ref{sect:standard_peeling}. The detailed implementation is as follows:
\begin{enumerate}
    % 1
    \item \textbf{Creating a temporary measurement set.} The target bright source model constructed in Step~\ref{item:step3} of $\S$~\ref{sect:standard_peeling} is 
    phase-centered on the source itself. To preserve its astrometric and photometric integrity, we added the target bright source model back \emph{after} applying the inverted gain and \emph{prior} to  restoring the original phase center in Step~\ref{item:step5}. We used the \texttt{split} task to extract the \texttt{CORRECTED\_DATA} column of \textbf{Temp2.ms} into a new temporary measurement set, \textbf{Temp\_restore.ms}, and adding the bright source model to the DATA column. % writing the data to its \texttt{DATA} column. 
    % 2
    \item \textbf{Adding back the target bright source model.} Following the same strategy used to restore the initial sky model in Step~\ref{item:step6}, we first executed the \texttt{tclean} task on the \textbf{Temp\_restore.ms} with \texttt{niter=0} to generate a structurally valid but empty \texttt{MODEL\_DATA} column. We then copied the \texttt{MODEL\_DATA} column of \textbf{Bright\_source\_model.ms}, which contains the best-fitting model of the target bright source, into the corresponding \texttt{MODEL\_DATA} column of \textbf{Temp\_restore.ms}. Finally, we used the \texttt{uvsub} task with \texttt{reverse=True} to added the target bright source model back to the visibilities, writing the result directly to the \texttt{CORRECTED\_DATA} column of \textbf{Temp\_restore.ms}. 
    % 3
    \item \textbf{Restoring the original phase center and producing the final science dataset.} We next restored the original phase center by applying the \texttt{phaseshift} task to all data columns of \textbf{Temp\_restore.ms}, producing a new measurement set, \textbf{Temp3.ms}. The \texttt{CORRECTED\_DATA} column of \textbf{Temp3.ms} was then copied into the \texttt{DATA} column of \textbf{Final\_peeling.ms} using the \texttt{split} task. Finally, we performed Step~\ref{item:step6} to add back the initial sky model, yielding the \texttt{CORRECTED\_DATA} column of the final \textbf{Final\_peeling.ms}. In this dataset, the target bright source within the \texttt{peeling} mask region is direction-dependently calibrated, while all other emission, including sources outside the mask, retains the original direction-independent calibration state, with the original phase center preserved for the entire dataset.
\end{enumerate}

\begin{figure*}%[htbp]
    \begin{minipage}[t]{0.999\linewidth}
    \centering
    \includegraphics[width=0.95\textwidth,angle=0,scale=1]{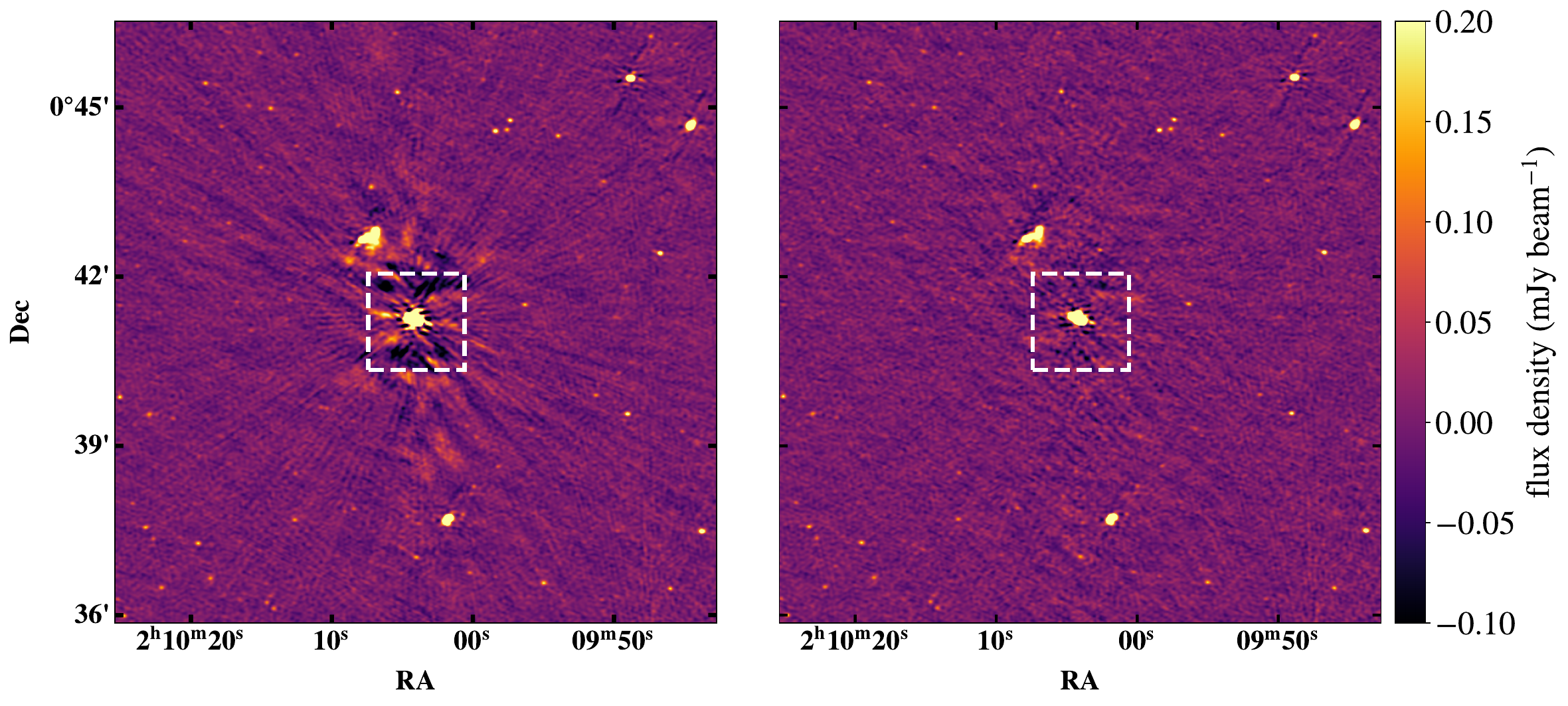}
    \caption{Comparison of the effects of applying the optimized \texttt{peeling} with model restoration to a bright source in the J0210-1 field. The white dashed box outlines the mask region. The left panel shows the image after the final round of self-calibration, where prominent artifacts around the bright source hinder the detection and measurement of neighboring sources. The right panel presents the image after applying the optimized \texttt{peeling} procedure, where artifacts associated with the bright source are effectively suppressed, producing a significantly flattened background and substantially enhancing the clarity and visibility of nearby sources, while the intrinsic morphology of the bright source itself remains well preserved.}
    \label{fig:2_peelingl}
    \end{minipage}
\end{figure*}

Fig.~\ref{fig:2_peelingl} illustrates the effectiveness of our optimized \texttt{peeling} procedure. In the right panel, artifacts surrounding the bright source are strongly suppressed, resulting in a significantly flatter background over the affected region, while the overall flux density of the source is largely preserved without obvious structural distortion. However, this demonstrates that the adopted model-restoration strategy successfully maintains the fidelity of the target source emission while substantially mitigating imaging artifacts arising from direction-dependent effects, its limitations are discussed in $\S$~\ref{sect:4.2}.

%%%%#### Figure 7 #####

\subsection{Global impact of peeling}
\label{sect:global_impact}

\begin{figure*}%[ht!]
    \begin{minipage}[t]{0.999\linewidth}  %% use \textwidth for full-page width crossing two columns
    \centering
    \includegraphics[width=0.9\textwidth,angle=0,scale=1]{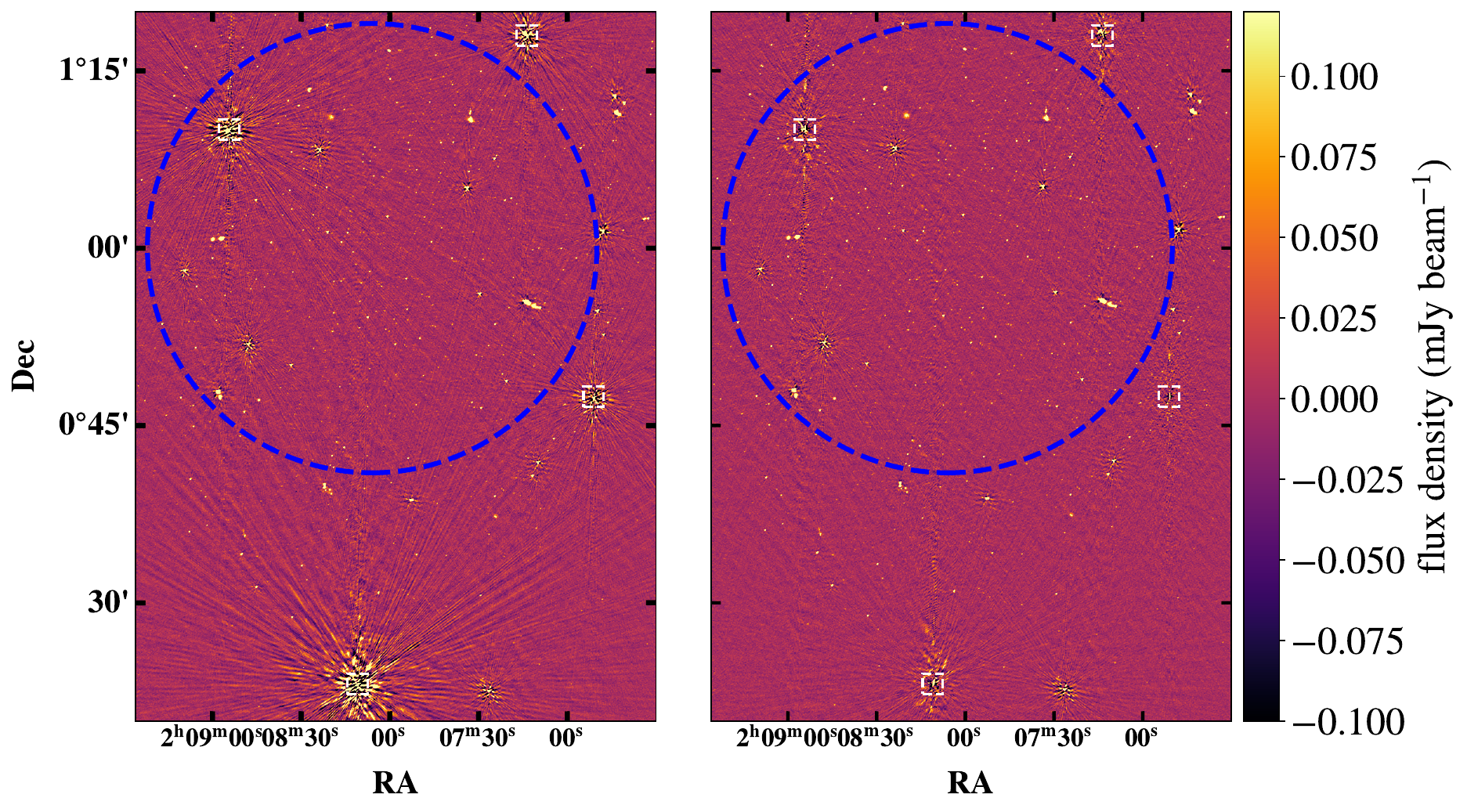}
    \caption{Comparison of the J0210-3 subfield images before (left) and after (right) the \texttt{peeling} procedure. The white dashed boxes outline the \texttt{peeling} mask regions of the four bright sources. The blue circle represents the uGMRT Band\,4 primary beam (PB), defining the main science region. The brightest source lies well outside the PB but still induces significant direction-dependent effects within it. This plot therefore extends the field of view beyond the nominal PB to illustrate its impact.}
    \label{fig:3+4_peeling_all}
    \end{minipage}
\end{figure*}

The J0210 field contains several bright radio sources, including four particularly strong sources in the J0210-3 subfield, each with total flux densities greater than 35\,mJy, where 35\,mJy corresponds to the faintest source among them. Artifacts associated with these sources significantly degrade the image quality across the field of view. To mitigate these effects, the peeling procedure was applied sequentially to the four targeted sources in order of decreasing flux density.Fig.\ref{fig:3+4_peeling_all} presents a comparison of the entire J0210-3 subfield before and after the peeling procedure, visually illustrating the improvement in image quality after peeling.

%%%%#### Figure 7 #####

\begin{figure}[ht!]
    \begin{minipage}[t]{0.999\linewidth}  %% use \textwidth for full-page width crossing two columns
    \centering
    \includegraphics[width=0.99\textwidth,angle=0,scale=1]{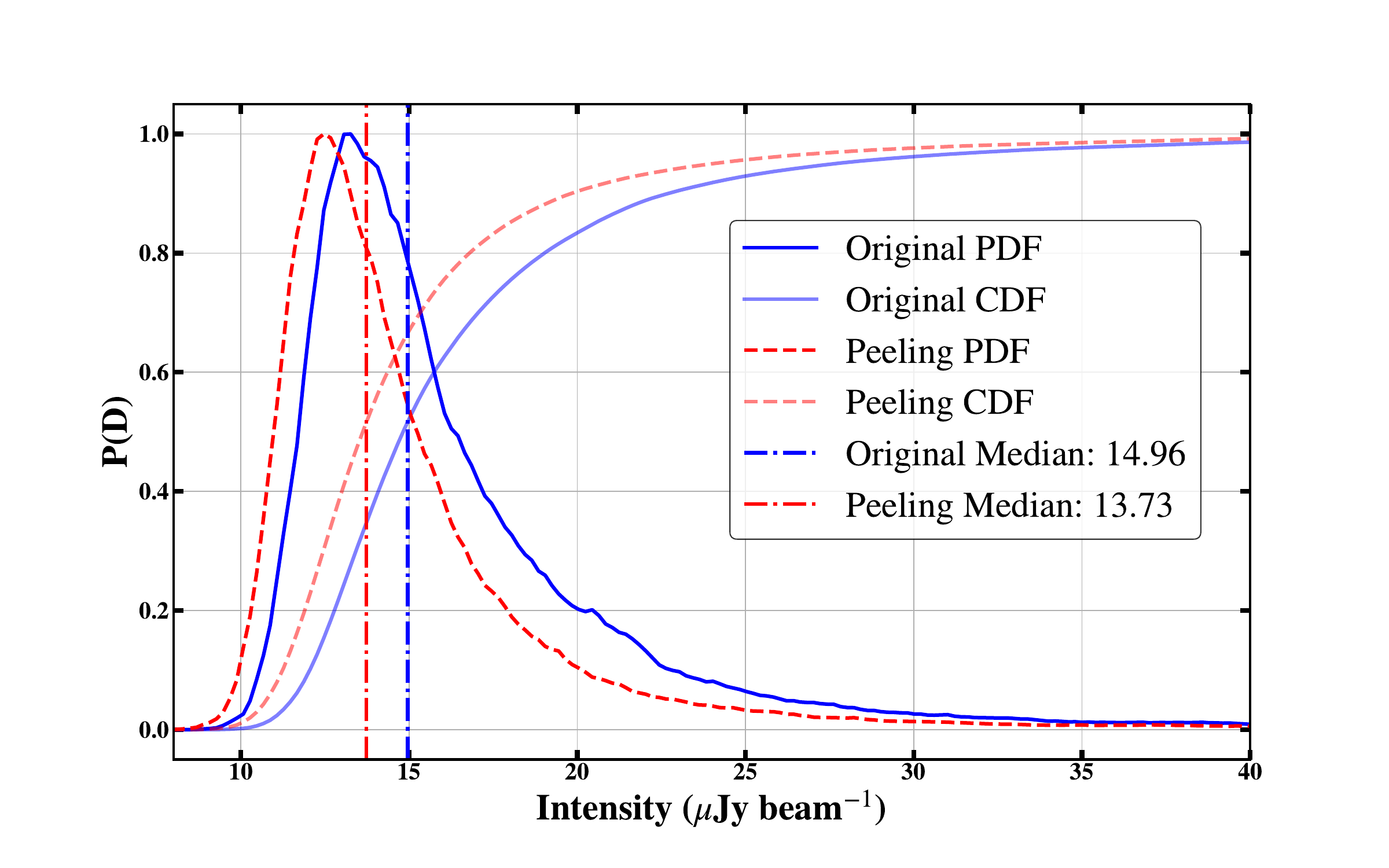} %.png/jpg/pdf/eps extension is not required
    \caption{Comparison of background RMS distributions in the J0210-3 subfield before (blue) and after (red) sequential \texttt{peeling} of the four bright sources. Solid lines show the normalized probability density function (PDF) of the RMS values from the \texttt{PyBDSF} RMS map, while light lines show the corresponding cumulative distribution functions (CDFs). Vertical dash-dotted lines indicate the medianRMS values, which decrease from  14.96\,$\mu$Jy/beam to 13.73\,$\mu$Jy/beam after \texttt{peeling}. The PDF peaks, representing the most probable RMS values, shift from 13.26\,$\mu$Jy/beam to 12.47\,$\mu$Jy/beam.} 
    \label{fig:3+4_rms}
    \end{minipage}
\end{figure}

To quantify the improvement in sensitivity achieved through \texttt{peeling}, we analyzed the 
background RMS map produced during the preprocessing step of \texttt{PyBDSF}\textsuperscript{\cite{Mohan2015}}. Fig.~\ref{fig:3+4_rms} presents the normalized probability density function (PDF) and cumulative distribution of the background RMS values in the J0210-3 subfield before and after the sequential \texttt{peeling} of the four bright sources. The \texttt{peeling} procedure systematically reduces the residual image noise: the median RMS decreases from 14.96\,$\mu$Jy/beam to 13.73\,$\mu$Jy/beam, and the peak of the PDF, which corresponds to the most probable RMS value, shifts from 13.26\,$\mu$Jy/beam to 12.47\,$\mu$Jy/beam. This reduction reflects the effective suppression of direction-dependent artifacts, resulting in a globally flatter background and an enhancement of overall image fidelity and sensitivity. As a direct consequence, the detectability of faint sources is substantially improved. After sequentially peeling the four bright sources, the number of sources detected above the $3\,\sigma$ threshold in the vicinity of the bright source (within 10\,arcmin) shown in Fig.~\ref{fig:3+4_peeling} increases from 154 to 183. Across the entire image, the total number of detected sources increases from 3260 to 3474, corresponding to an additional 214 sources. % with an additional 183 sources detected above the $3\,\sigma$ threshold following the full sequential \texttt{peeling} procedure.

%%%%#### Figure 8 #####

\section{Discussion}
\label{sect:discussion}

We develop a Python-based, CASA-compatible \texttt{peeling} framework and demonstrate its effectiveness using uGMRT Band\,4 imaging data. The framework mitigates direction-dependent artifacts, reduces image background noise, and enables both the subtraction and preservation of bright target sources. While the framework significantly improves image fidelity and sensitivity, a key challenge inherent to any \texttt{peeling}-based approach remains accurately modeling the bright source. The fidelity of this model ultimately limits the effectiveness of direction-dependent calibration and subsequent source subtraction. Furthermore, as discussed in $\S$~\ref{sect:4.2}, the restoration of bright sources is inherently limited, and the recovered flux densities and morphologies should be interpreted with caution, particularly for high-precision analyses. 

\subsection{Challenges in modeling target bright sources}\label{sect:4.1}
As described in $\S$~\ref{sect:standard_peeling}, modeling the target bright source requires careful control of the deconvolution depth. Excessively deep deconvolution within a finite \texttt{peeling} mask can lead to over-cleaning, whereby emission from faint sources adjacent to the bright target is inadvertently incorporated into the model. These components are then partially or fully removed during model subtraction, compromising the flux integrity of the neighboring faint sources. Even when such components are reintroduced through the model-restoration strategy, they may introduce new local direction-dependent artifacts, thereby degrading the surrounding image quality. 

Conversely, under-cleaning, i.e., insufficiently deep deconvolution, fails to capture the majority flux of the target bright source, leaving substantial residual emission after subtraction. This limits the suppression of direction-dependent artifacts and reduces the overall effectiveness of the \texttt{peeling} process. Consequently, modeling the target bright source requires a careful balance between over- and under-cleaning, guided by close inspection of the residual image.

The ideal scenario is one in which the \texttt{peeling} mask contains only the target bright source, as illustrated in Fig.~\ref{fig:2_peelingl}. In such cases, deeper deconvolution can be safely applied, enabling the model to capture nearly all of the source flux without incorporating extraneous emission, thereby yielding a highly accurate representation. In practice, however, the \texttt{peeling} mask often encompasses both the target bright source and nearby faint sources (e.g., Fig.~\ref{fig:3+4_peeling}). Under this condition, deep deconvolution risks absorbing flux from neighboring sources into the bright-source model.

A pragmatic compromise is therefore to construct a model that captures the bulk of the target bright source emission while allowing a small level of residual emission to remain. Such residuals can be reliably excluded during subsequent source detection following the peeling process. Overall, robust modeling of bright target sources relies on iterative residual monitoring during deconvolution, ensuring that the dominant emission is captured while minimizing adverse effects on neighboring faint sources.

\subsection{Limitations of restoring bright sources}\label{sect:4.2}
The challenges in accurately modeling target bright sources also highlight intrinsic limitations in restoring these sources after the \texttt{peeling} process. In principle, reliable restoration is only achievable when the \texttt{peeling} mask contains exclusively the target bright source. When the mask includes nearby faint emission, particularly for extended sources whose structure may be entangled with artifacts, the model can be contaminated, leading to biased flux restoration. 

Even when the bright source appears well modeled, the restoration process introduces additional limitations. The restored model, constructed from idealized, noiseless, clean components, replaces the original data within the masked region. Consequently, it lacks the true localized thermal noise, unresolved emission, and baseline-dependent residual errors present in the original visibilities. As a result, both photometry and morphology of restored sources should be treated with caution. While the total flux density may be approximately preserved, fine structural details and realistic noise characteristics are not guaranteed to be accurately recovered. 

A rigorous validation would involve injecting simulated sources with known properties into realistic visibility data and processing them through the full pipeline to assess flux recovery, astrometry, and uncertainties. Although we performed preliminary tests, the dominant direction-dependent effects in our data, primarily from the antenna primary beam and ionospheric/tropospheric variations, are difficult to simulate with high fidelity, limiting the realism of such tests. 

To isolate systematic uncertainties introduced by the model-restoration step itself, apart from the direction-dependent gain calibration, we performed a validation test using real sources in the J0210-1 field. We selected two isolated point sources and two moderately extended sources that (i) do not require peeling, (ii) are only weakly affected by direction-dependent effects, and (iii) have high signal-to-noise ratios. These sources were then processed through the same ``model extraction + model re-injection'' procedure as adopted in the optimized peeling workflow, but with the direction-dependent gain calibration step intentionally omitted. We quantitatively compared the recovered flux densities, source sizes, centroids, and local background RMS before and after the model-restoration procedure. The results are summarized in Table~\ref{tab:restoration_validation}, which lists the right ascension (RA), declination (DEC), flux density, source size (major and minor axes), and the median local RMS measured within a $200''\times200''$ region centered on each source. We find no evidence for systematic offsets in any of these quantities within the measurement uncertainties. This indicates that the model-restoration step itself does not introduce significant biases in flux density, morphology, astrometry, or noise properties, at least for sources that are not strongly affected by direction-dependent effects.

However, this test does not provide a complete uncertainty budget for the full peeling procedure; it only constrains the lower limit of the systematic uncertainties associated with the model-restoration step. We thus recommend caution when using restored flux densities and morphologies, particularly for analyses requiring high-precision photometry or detailed structural interpretation.

%\textcolor{cyan}{The results are shown in Table\ref{tab:restoration_validation} , which lists the right ascension (RA), declination (DEC), flux density, source size (Maj, Min), and local RMS median within a $200''\times200''$ region centered on each source for the four sources (two point sources and two extended sources) before and after peeling. None of these quantities exhibit systematic increases or decreases. This confirms that the model-restoration step itself does not introduce significant systematic biases in flux, morphology, astrometry, or noise properties.}

\begin{table*}[h!]
\centering
\caption{Validation test results for the model-restoration step (without direction-dependent gain calibration).}
\label{tab:restoration_validation}
\begin{tabular}{@{} c c c c c c c c @{}}
\toprule
Source & Epoch & RA (\si{\degree}) & Dec (\si{\degree}) & Flux density (mJy) & Maj (\arcsec) & Min (\arcsec) & RMS median (\si{\mu Jy\,beam^{-1}}) \\
\midrule
\multirow{2}{*}{point source-1} & before & 32.9490631 & 1.1758253 & 0.942 & 4.487 & 3.589 & 10.92 \\
                                 & after  & 32.9490637 & 1.1758247 & 0.941 & 4.490 & 3.579 & 10.93 \\
\addlinespace
\multirow{2}{*}{point source-2} & before & 33.0246101 & 0.4580520 & 0.529 & 5.031 & 3.849 & 11.35 \\
                                 & after  & 33.0246082 & 0.4580541 & 0.533 & 5.089 & 3.853 & 11.35 \\
\addlinespace
\multirow{2}{*}{extended source-1} & before & 33.0318419 & 0.4337568 & 0.975 & 5.698 & 3.648 & 11.91 \\
                                   & after  & 33.0318431 & 0.4337584 & 0.973 & 5.694 & 3.621 & 11.92 \\
\addlinespace
\multirow{2}{*}{extended source-2} & before & 33.1701226 & 1.2238597 & 0.577 & 10.154 & 4.410 & 12.51 \\
                                 & after  & 33.1701234 & 1.2238600 & 0.575 & 10.207 & 4.385 & 12.53 \\
\bottomrule
\end{tabular}
\end{table*}

\section{Conclusions}
\label{sect:conclusions}

Based on our results and analyses presented in $\S$~\ref{sect:peeling}, we draw the following conclusions: 

\begin{enumerate}
    \item The \texttt{peeling} framework developed in this work effectively subtracts the bright-source models and suppresses their associated direction-dependent artifacts in uGMRT 650\,MHz imaging data. This leads to a significantly flatter background over extended regions surrounding the peeled sources, thereby improving overall image fidelity and enhancing both the detectability and the flux density accuracy of neighboring faint sources. 
    \item Our optimized ``model restoration" strategy preserves the overall flux densities of scientifically relevant bright sources while effectively removing associated direction-dependent artifacts. This capability overcomes a key limitation of standard subtraction-only \texttt{peeling} approaches, enabling bright sources to be retained when they themselves are science targets.
   \item For fields containing multiple bright radio sources, sequential \texttt{peeling} systematically reduces background noise across the field of view. In the J0210-3 subfield, the median background RMS decreases by $\sim1.2\,\mu$Jy\,beam$^{-1}$ after sequential \texttt{peeling}, leading to improved image sensitivity and, consequently, the detection of an additional 214 sources above the $3\,\sigma$.
    \item The proposed \texttt{peeling} framework is modular, automated, and fully compatible with standard CASA-based calibration and imaging pipelines. While the framework has been rigorously validated using uGMRT Band4 imaging data, its architecture enables broad adaptation to other interferometric datasets. The full implementation (Software Availability: https://zenodo.org/records/18399370) is publicly released with this paper. 
\end{enumerate}

\section{ACKNOWLEDGEMENTS}
FXA acknowledges the support from the National Natural Science Foundation of China (12303016) and the Natural Science Foundation of Jiangsu Province (BK20242115). XZZ ackonwledges the support from the National Key Research and Development Program of China (2023YFA1608100),  the National Science Foundation of China (12233005, 12173088), the China Manned Space Program with grants nos. CMS-CSST-2025-A08 and CMS-CSST-2025-A20, and in part by Office of Science and Technology, Shanghai Municipal Government (grant Nos. 24DX1400100, ZJ2023-ZD-001). 
%%Example wide rotated table =========================================== 
\newpage

\normalsize
%%%%%%%%%%%%%%%%%%%% REFERENCES %%%%%%%%%%%%%%%%%%
%% The best way to enter references is to use BibTeX:
%% How to copy an entry of reference: 
%%    (1) open ADS abstract page of a paper, e.g. https://ui.adsabs.harvard.edu/abs/2006ApJ...642..868H/abstract
%%    (2) click lower left tab "Export Citation"; (3) click "Copy to Clipboard"; (4) Paste it to your_Ref.bib file
\clearpage

\bibliographystyle{ati} %% Bibliography Style File -- ATI references formating file ati.bst
\bibliography{ati}      %% ati.bib or your_Ref.bib: list of BibTeX entries of references from ADS, etc., can contain literature not actually cited in this paper

\label{lastpage}

\end{document}